\documentclass[prd,twocolumn,aps,amsmath,nofootinbib,superscriptaddress]
{revtex4}
\usepackage{graphicx}
\usepackage{bm}
\usepackage{epsfig}

\def\ltap{\;\centeron{\raise.35ex\hbox{$<$}}{\lower.65ex\hbox{$\sim$}}\;}
\def\gtap{\;\centeron{\raise.35ex\hbox{$>$}}{\lower.65ex\hbox{$\sim$}}\;}

\begin{document}


\preprint{\vbox{\hbox{68-2532} \hbox{hep-ph/0412007}  }}

\title{Light Scalars and the Generation of Density Perturbations 
During Preheating or Inflaton Decay}

\author{Lotty Ackerman}
\affiliation{California Institute of Technology, Pasadena, CA 91125}
\author{Christian W.~Bauer}
\affiliation{California Institute of Technology, Pasadena, CA 91125}
\author{Michael L.~Graesser}
\affiliation{California Institute of Technology, Pasadena, CA 91125}
\author{Mark B.~Wise}
\affiliation{California Institute of Technology, Pasadena, CA 91125}

\begin{abstract}
Reheating after inflation can occur 
through inflaton decay or 
efficient parametric resonant production of 
particles from the oscillation of the inflaton. 
If the particles produced 
interact with scalars that were light during inflation, then significant 
super-horizon density perturbations are generated during this era. 
These perturbations can be highly non-Gaussian. 
\end{abstract}

\maketitle

Measurements of the cosmic microwave background
radiation~\cite{cobe,wmap} have clearly
shown the presence of super-horizon primordial density fluctuations at
roughly one part in $10^5$.
Inflation provides a natural explanation for such density fluctuations,
since vacuum fluctuations
of the inflaton (or any other light scalar field) get pushed outside of
the horizon and enter
at a much later time as classical density
perturbations~\cite{inflation_perturbation}.
Recently, Dvali, Gruzinov and Zaldarriaga \cite{DGZ,DGZ2} and 
Kofman \cite{kofmanoriginal} (DGZK) have shown in a 
number of scenarios how the interactions of such additional light 
fields to, e.g. the  
inflaton, could also   
generate 
adiabatic
density fluctuations, 
independent of those created by the inflaton 
dynamics. In this scenario 
the size of non-Gaussian perturbations can 
be much larger than what 
occurs in single-field inflationary models~\cite{DGZ2,Zaldarr}.

This is achieved by coupling a light scalar field 
to a heavier field that at some time subsequent to inflation dominates 
the energy of the Universe, such that the particle properties 
of this heavier field are modified by the fluctuations of 
the light field. 
When the heavier particle decays, spatial fluctuations in either 
its mass or its decay width generate energy density 
perturbations in the radiation.   
This is because before 
reheating the universe
is matter dominated, with the oscillating heavier particle dominating the total
energy density,
while after the decay the universe is radiation dominated. As the energy
density in matter redshifts slower than energy density in radiation, 
regions of the universe where the decay occurs at a later time 
stay matter dominated longer and will be denser than regions
where decay happens earlier. This gives density perturbations of order
\begin{eqnarray}\label{deltarhosimple}
\frac{\delta\rho}{\rho} \sim  -\frac{\delta \Gamma}{\Gamma} \sim
\frac{\delta  \tau}{
\tau}\,,
\end{eqnarray}
where
\begin{eqnarray}
\tau \equiv t_{\rm RH} - t_0
\end{eqnarray}
is the time between the end of inflation $(t_0)$ 
and reheating $(t_{\rm RH})$. The 
evolution of density perturbations in 
this scenario
has been studied in
detail in \cite{DGZ,addpert}.

In a similar way, modifications to the particle properties 
of the particles {\em produced} during reheating can also introduce 
energy density perturbations. 
Density perturbations are created if the decay 
products interact with 
fields that 
were light during inflation. 

To see this, we need to discuss how the inflaton reheats. 
Suppose reheating occurs through direct (Born) decay of the inflaton. 
Then a fluctuation in the mass of the decay product $\chi$ modifies 
the inflaton decay width, because
of the dependence of the available phase space on the masses of the
final state particles. These lead to {\em calculable} density 
perturbations 
since the exact dependence of the width on the mass of the
light particles can be computed in any given model \cite{Zaldarr,Postma}. 
If, for example,
the inflaton decays via $\Phi \rightarrow \chi \chi $ 
then the tree-level decay width is modified from 
phase space by an amount
\begin{equation}
\frac{\delta \Gamma}{\Gamma} = -2 \frac{\delta m^2_{\chi}}{m^2_{\phi}
-4 m^2_{\chi}}
+ 2 \left(\frac{\delta m^2_{\chi}}{m^2_{\phi}-4 m^2_{\chi}} \right)^2 + \cdots\,.
\end{equation}
If $\Phi$ decays  
near threshold, then the resulting 
density perturbation dependence on $\delta m^2_{\chi}$ 
can be large
and highly non-linear. 

\begin{figure}[t!]
  \centerline{\mbox{\epsfysize=2.5truecm 
\hbox{\epsfbox{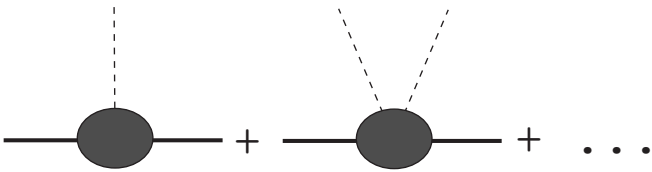}} } }
  {\caption[1]{Quantum corrections may generate a 
dependence of $m^2_{\chi}$ on super-horizon 
fluctuations in $\sigma$.}
\label{fig:feynmandiagram} }
\end{figure}

We expect a $\delta m^2_{\chi}$ with a super-horizon spatial 
variation to be generated if $\chi$ interacts with a field 
$\sigma$ that was light during the inflationary era and 
through to the era of reheating. 
Note that even in the absence of direct 
couplings of the fields $\chi$ and $\sigma$, they 
are expected to interact indirectly through some 
intermediate states. Quantum corrections will typically generate 
a dependence of $m^2_{\chi}$ on the super-horizon 
fluctuations of $\sigma$ 
at some order in perturbation theory, as indicated by Fig.~\ref{fig:feynmandiagram}. In this paper we focus mainly  
on the effect that fluctuations in the mass of the particles produced
during reheating or preheating have on density perturbations. 
 
Besides reheating through 
direct Born decay, the inflaton may instead reheat the universe 
through parametric resonance 
(preheating) 
\cite{parametric,STB,preheatingstudy,preheatingstudy2,preheatingstudy3}.
Preheating can be very efficient and
be completed very soon after inflation, within
${\cal O}(10-100)$ oscillations of the inflaton field about its minimum.
Whether this process of reheating dominates over the Born 
decay into bosons or fermions depends on the parameters 
of the model \footnote{The growth of 
perturbations  
during the matter-dominated era of the oscillating 
inflaton 
has been studied 
in \cite{infloscpert} and, if parametric resonance 
occurs,  
in \cite{preheatpert}.}.

In the scenario of DGZK, additional 
density perturbations can be created during preheating, 
by modifying the time it takes for parametric 
resonance to complete and for the universe to thermalize. 
The size of this time interval depends on the parameters 
of the model, and in particular on the mass of the produced particles, 
which we discuss below in a simple model. 
This is the main subject of this paper. 
Depending on how efficient preheating is, the size of 
the time interval can have a weak or 
strong sensitivity to the mass 
of the decay products. 

We use the canonical model of preheating 
and add a scalar $\sigma$ which we 
assume is light during inflation so that it acquires super-horizon 
perturbations $\delta \sigma(x) \sim H_{\rm inf}$ 
during that era. For this to occur it is necessary that during 
inflation its mass 
satisfies $m_{\sigma} < H_{\rm inf}$. 
$\sigma$ is assumed to interact 
more strongly with the $\chi$ compared to $\Phi$. 
The interactions we consider are  
\begin{eqnarray}
-{\cal L}_I = \frac{g^2}{2} \Phi^2 \chi^2
+ \mu \, \chi^2 \sigma + \frac{\lambda}{2} 
\chi^2 \sigma^2 +\frac{m^2_{\chi}}{2}
\chi^2 + \frac{m^2_{\sigma}}{2} \sigma^2.
\label{interactions}
\end{eqnarray}
A $Z_2$ symmetry $\chi \rightarrow - \chi$ has been 
imposed for simplicity. Self-interactions $\sigma^4$ and 
$\chi^4$ are assumed to be irrelevant during the first 
stage of preheating defined below.  
We assume that at the end of inflation 
the fields $\chi$ and $\sigma$ are near enough to the minimum 
of their potential
so that we can neglect the motion of their zero modes.

Inflation ends when $t=t_0 \simeq 1/m_{\Phi} $ 
and is followed by a matter-dominated
era described by rapid oscillations of the inflaton about the minimum of 
its potential which we assume to be 
\begin{equation}
V(\Phi) =\frac12 m^2 _{\Phi} \Phi^2 ~.
\end{equation}
For simplicity we assume that the inflationary potential is also 
described by this simple quadratic form, 
giving rise to chaotic inflation~\cite{chaotic_inflation}. During 
inflation, $H_{\rm inf} \simeq m_{\Phi}$. 
At the end of inflation $\Phi =\Phi_0 \simeq m_{\rm pl}/3$ and 
thereafter decays as $\Phi(t) \simeq m_{\rm pl}/3 m_{\Phi} t$.

For large enough coupling $g$, 
these oscillations trigger parametric resonance, and the energy density in 
$\chi$ increases exponentially~\cite{parametric}. If this process is
efficient, the universe eventually 
is dominated by the $\chi$ particles, which
then thermalize the universe at some later time through 
its interactions with Standard Model or Grand Unified 
Model particles.

The  
perturbations in the inflaton give rise to adiabatic density
perturbations, whose size depend on the form of the inflaton potential.
In this letter we concentrate on the density perturbations generated
from the fluctuations in the $\sigma$ scalar field. In de Sitter space~\cite{LindeBook}
\begin{eqnarray}
\langle \sigma^2(0)\rangle = \frac{H^2_{\rm inf}}{4 \pi^2}N\,,\label{delsig1}\\
\langle \sigma(x) \, \sigma(y) \rangle = \frac{H^2_{\rm inf}}{4 \pi^2}\label{delsig2}\,,
\end{eqnarray}
where in Eq.~(\ref{delsig1}) $N$ is the number of e-foldings during inflation. In Eq.~(\ref{delsig2})
the comoving coordinates $x$, $y$ are well seperated and we neglect the logarithmic dependence on $|x-y|$.

The 
$\chi$ field does not acquire super-horizon perturbations because 
its effective mass 
\begin{eqnarray}\label{mchieff}
m_{\chi, \rm eff}^2=m_\chi^2 + g^2 \,|\Phi|^2 + \lambda \frac{H^2_{\rm inf}}{4\pi^2} N
\end{eqnarray}
during inflation is 
larger than the Hubble parameter 
for parameter values which allow for efficient parametric
resonance. 
Henceforth we absorb the $\lambda H^2_{\rm inf}N/(4 \pi^2)$ into $m_\chi^2$. Treating $\sigma$ as an external field, its 
fluctuations can be absorbed into fluctuations in the mass of
the field $\chi$, 
\begin{eqnarray}\label{massfluctuations}
\delta m_\chi^2 &=& 2 \, \mu \, \delta \sigma + \lambda \,
\delta \sigma ^2.
\end{eqnarray}
where we have used
\begin{eqnarray}
\delta \sigma \equiv \sigma - \langle \sigma \rangle \,, \quad 
\delta \sigma^2 \equiv \sigma^2 - \langle \sigma^2 \rangle\,,
\end{eqnarray}
and we will impose $\langle \sigma \rangle=0$. 
The size of the fluctuations $\delta m_\chi^2$ is determined by the two-point function
\begin{eqnarray}
\langle \delta m_\chi^2(x)\, \delta m_\chi^2(y) \rangle = 4 \mu^2 \langle \sigma(x) \, \sigma(y) \rangle + 
2 \lambda^2 \langle \sigma(x) \sigma(y) \rangle^2\nonumber
\end{eqnarray}
Using Eqs.~(\ref{delsig1}) and (\ref{delsig2}) we find the fluctuations for widely separated comoving coordinates $x$ and $y$ to be of order
\begin{eqnarray}
\delta m_\chi^2 \sim \sqrt{\mu^2 H^2_{\rm inf} + \lambda^2 H^4_{\rm inf}}\,.
\end{eqnarray} 

While the field $\sigma(x)$ is Gaussian, the fluctuation $\delta m_\chi^2$ is only Gaussian for $\lambda=0$. For $\lambda H^2_{\rm inf} \gg \mu H_{\rm inf}$,  $\delta m_\chi^2$ is highly non-Gaussian.  For example, consider in this limit the three-point function for equally separated comoving coordinates. 
One finds  for the analog of skewness
\begin{eqnarray}
&&\frac{\langle \delta m_\chi^2(x) \delta m_\chi^2(y) \delta m_\chi^2(z) \rangle}{ \langle \delta m_\chi^2(x) \delta m_\chi^2(y) \rangle ^{3/2}} \\
&& \hspace{1cm}
= \frac{8 \langle \sigma(x) \sigma(y) \rangle \langle \sigma(y) \sigma(z) \rangle \langle \sigma(z) \sigma(x) \rangle }{2^{3/2} \langle \sigma(x) \sigma(y) \rangle ^3}
= 2 \sqrt{2}\nonumber\,.
\end{eqnarray}
For the remainder of this paper we set $\mu=0$ which corresponds 
to imposing a $\sigma \rightarrow -\sigma $ symmetry. We make 
this decision to simplify the analysis of the backreaction of 
$\chi$ on $\sigma$ discussed below. Then 
\begin{equation} 
\frac{\delta m^2_{\chi}}{m^2_{\Phi}} \sim \lambda 
\frac{H^2_{\rm inf}}{m^2_{\Phi}} \simeq \lambda ~.
\end{equation} 
These fluctuations in $\delta m^2_{\chi}$ are non-Gaussian 
and always 
positive.  

This situation would be 
excluded if this were the only source of density perturbations. 
A more 
interesting scenario in this situation would be if the dominant source 
of perturbations came from the inflaton potential. Then the perturbations 
generated during preheating providing a sub-dominant, non-Gaussian 
contribution. Since here the source for the non-Gaussian perturbations 
is {\em not} the same as the source -the inflaton- 
providing the dominant Gaussian contribution, 
the current limits on non-Gaussianity~\cite{WMAPGaussian}
do not apply, since those limits assume that the non-Gaussian and 
Gaussian perturbations are generated by the same field. 
 
 We define
preheating to last until significant particle production of 
$\chi$ occurs and the energy densities in $\Phi$ and 
$\chi$ become equal. The duration of this stage 
depends on $m_{\chi}$ and coupling constant
$g$, 
\begin{eqnarray}\label{tRH}
\tau = \tau(g,m_\chi)~.
\end{eqnarray}
Fluctuations in $m_\chi$ and the coupling $g$ give rise to density
fluctuations from Eq.~(\ref{deltarhosimple}). 

Fluctuations in $g$ can be generated if 
it is replaced by an 
effective coupling 
\begin{equation} 
g^2_{\rm eff}=g^2 \left(1 + \frac{ \sigma^2}{M^2} \right) ~,
\label{geff}
\end{equation} 
where $M$ is some mass scale \cite{Kofman}. The $\sigma$ 
dependence of $g_{\rm eff}$ generates 
non-Gaussian perturbations $\delta_g \equiv 
\delta g^2/g^2 = H ^2_{\rm inf}/M^2$. It also  
modifies the large time-dependent mass of $\chi$, an 
effect that is distinct 
from modifying $m_{\chi}$.  
Here too we have to worry about the 
backreaction of $\chi$ on $\sigma$.

Next we describe our numerical method 
for determining the energy density in 
$\chi$ during preheating. 
Neglecting the backreaction of $\chi$ on the inflaton, 
which only becomes significant 
at the end of the preheating stage when $\rho_{\chi} = 
\rho_{\Phi}$ \cite{preheatingstudy}, the 
equation of motion for the fields $\chi\equiv \hat{\chi}(a_0/a)^{3/2}$ 
($a$ is the scale factor) and $\sigma$ are
\begin{eqnarray}
{\hat{\chi}}''_k + \left[A_k + 2 q \cos(2 (z-z_0))\right] \hat{\chi}_k = 0
\end{eqnarray}
\begin{equation} 
\delta \sigma '' + \frac2z \delta \sigma' 
+m^2_{\sigma ,\rm eff} \delta \sigma =0~,
\end{equation} 
where derivatives are with respect to
$z \equiv m_\Phi t\,$ and we have chosen $z_0 \equiv 1$. 
We have defined
\begin{eqnarray}
q &=& \frac{g^2 \Phi_0^2 a^3_0}{4 a^3m_\Phi^2} \equiv q_0 
\frac{a^3_0}{a^3} ~~,~~ a_0 \equiv a(t_0) \nonumber\\
A_k &=& \frac{1}{m^2_{\Phi}}
\left(k^2 \frac{a^2_0}{a^2}+ \widetilde m^2_{\chi} \right) + 2 q~,
\end{eqnarray}
and the mass parameters are given by
\begin{eqnarray} 
\widetilde m^2_{\chi}&=& m^2_{\chi} + \lambda \, \delta \sigma^2 \nonumber\\
m^2_{\sigma,\rm eff}&=&m^2_{\sigma} + \lambda \chi^2 \,. 
\end{eqnarray} 
The equation for $\hat{\chi}_k$ describes a time-dependent harmonic oscillator 
with frequency $\Omega^2_k = m_\Phi^2 \left[ A_k + 2 q \cos(2 (z-z_0)) 
\right]$. 
In the limit of a static universe and constant $\delta \sigma$ 
this equation reduces to the Mathieu equation.

Efficient parametric resonance requires $q_0 \gg 1$ and $\widetilde m_\chi \lesssim m_\phi$. Note that 
we included a term of order $\lambda H^2_{\rm inf} N$ into the definition of $m_\chi^2$, where $N$ is the number of e-foldings during inflation. The bound  $\widetilde m_\chi \lesssim m_\phi$ therefore  implies $N \lesssim \lambda^{-1}$. For the values of $\lambda$ we consider, this is a very weak bound on the number of e-foldings during inflation.

For a given value of $k$ the energy density in $\chi$ is 
\begin{eqnarray}\label{rhok}
\rho_k(z) = \Omega_k(z) N_k(z)~, 
\end{eqnarray}
where $N_k(t)$ is the number density for a mode with given wave number
$k$. The number density can be calculated by numerically 
solving for the Bogolyubov
coefficient, giving ~\cite{STB}
\begin{equation} 
N_k(t) =\frac{a^3_0}{2 \Omega_k a^3(t)} 
\left(\Omega^2_k(t)|\tilde{\chi}_k|^2 + m^2_{\Phi}
|\tilde{\chi}'_k|^2 \right)~,
\end{equation} 
with initial conditions $\tilde{\chi}_k(t_0)=1/\sqrt{2 \Omega_k}$, 
$m_{\Phi}\tilde{\chi}'_k(t_0) =- i \sqrt{\Omega_k/2}$.  
The field $\tilde{\chi}_k$ satisfies the same equation as 
$\hat{\chi}_k$ and is related to it (see Appendix B of \cite{STB}). 
The energy density is obtained by integrating
Eq.~(\ref{rhok})  
to obtain
\begin{eqnarray}
\rho_\chi(z) = \frac{1}{2 \pi^2} \int_0^\infty \!\!\! k^2 dk
\,\,\Omega_k(z) N_k(z)~.
\end{eqnarray}

The exponentially large 
number density of $\chi$ particles leads to a large backreaction on 
$\sigma$ that must be 
included to correctly determine the size of the 
effect we are describing. 
The backreaction of $\chi$ on $\sigma$ can have two effects: 
first, it can lead to production of large numbers of $\sigma$ particles, and 
second it gives rise to a large effective mass of the $\sigma$ field. 
The first effect was analyzed by Felder and Kofman~\cite{FK} using a
numerical lattice simulation of preheating and the
subsequent thermalization of the $\chi$ with the $\sigma$ fields. In their 
Figures 14 and 15 they show the number densities of 
$\Phi$, $\chi$ and $\sigma$. Their numerical 
results show that during preheating
the number 
density in $\sigma$ is much smaller than in either $\chi$ or 
$\Phi$ and its effect on the evolution of either $n_{\chi}$ or 
$n_{\Phi}$ is negligible. The second effect is more significant. 
Once $m_{\sigma,\rm eff}$ gets 
larger than $H$, the 
amplitude $\delta \sigma $ will decrease 
rapidly \cite{mazumdar}. To simplify the analysis 
we will assume that the 
dependence of $m_{\sigma, \rm eff}$ on $m_{\sigma}$ 
can be neglected.  To
estimate the time at which the backreaction becomes 
important, we compare the effective mass 
$m^2_{\sigma, \rm eff} \sim \lambda \langle\chi^2\rangle$ 
to the Hubble parameter. 
The ratio that determines their 
relative importance can be expressed as 
\begin{eqnarray}
\frac{m^2_{\sigma, \rm eff}}{3 H^2} = \frac{2\lambda}{3 g^2} 
\frac{\rho_{\chi}}{\rho_{\Phi}} \frac{m^2_{\Phi}}{H^2} ~,
\label{backreaction}
\end{eqnarray} 
where we have used $m_{\chi,\rm eff} \simeq g |\Phi|$ 
and $\rho_{\chi} \simeq g |\Phi| n_{\chi} 
\simeq g^2 \Phi^2 \langle\chi^2\rangle$~\cite{preheatingstudy}.
For $\lambda \sim 10^{-5}$ and $H\sim 2m_\Phi/300$ we find that 
this backreaction
becomes 
important when $\rho_\chi/\rho_\Phi \approx 3 \, g^2$. 
For $\lambda \sim 10^{-7}$ the backreaction becomes important 
when $\rho_\chi/\rho_\Phi \approx 300 \, g^2$.
In this letter we will 
not solve the full coupled set of differential equations, but rather deal with 
this backreaction by turning off $\delta m^2_\chi$ at the time $z_c$
when $m^2_{\sigma, \rm eff}=3 H^2$, i.e. defined implicitly by 
\begin{equation} 
\frac{\rho_{\chi}(z_c)}{\rho_{\Phi}(z_c)} \equiv \frac{g^2 }{\lambda} 
\frac{2}{3 z_c^2}~.
\label{backreaction2} 
\end{equation}
Although for different values of $\delta m^2_{\chi}$
the intercept time $z_c$ is different, that difference 
is second order in $\delta m^2_{\chi}$. 
It is then 
sufficient to use the $z_c$ obtained  
by setting $ \delta m^2_{\chi}=0$. If 
Eq.(\ref{backreaction2}) intercepts 
$R$ along a plateau corresponding to no particle production, 
then we make the conservative 
choice of cutting off the mass 
fluctuation at the location of the first intercept.

\begin{figure}[t!]
  \centerline{\mbox{\epsfysize=4.5truecm 
\hbox{\epsfbox{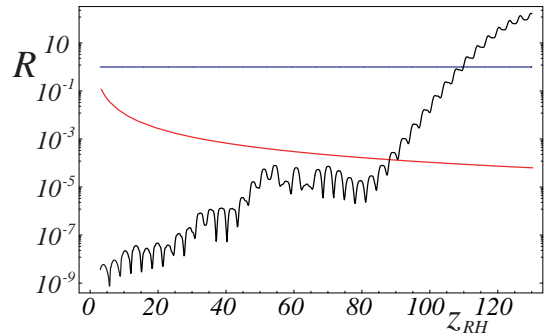}} } }
  {\caption[1]{Logarithmic plot of $R = \rho_\chi/\rho_\Phi$. The chosen 
parameters
   are $g = 4 \times 10^{-4}$, $m_{\Phi} = 10^{-6} m_{\rm pl}$,
$m^2_{\chi}/m^2_\Phi = 0.1$. 
Also shown is Eq.~(\ref{backreaction2}) with $\lambda=10^{-7}$.}
\label{fig:plot1} }
\end{figure}

\begin{figure}[t!]
  \centerline{ \mbox{\epsfxsize=9.5truecm \hbox{\epsfbox{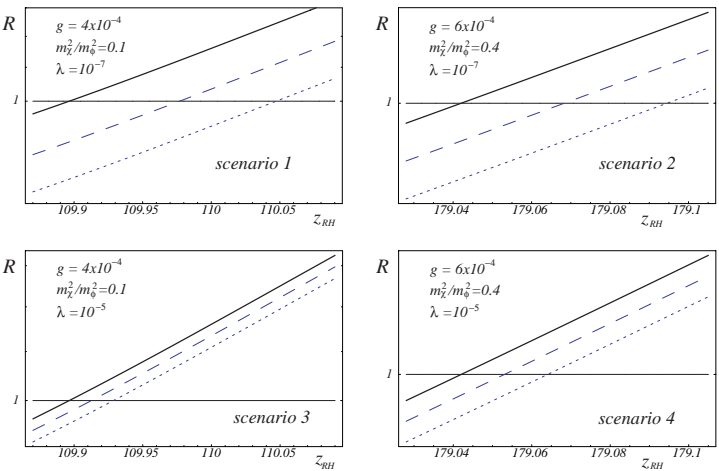}} } }
  {\caption[1]{Logarithmic plot of the effect of the mass fluctuation 
$\delta m_\chi^2$ on $z_{\rm RH}$. The chosen parameters
   are given in each figure. The solid line corresponds  to $\delta m^2_\chi = 0$, while the long and short dashed
lines correspond to
   $\delta m^2_\chi/ m^2_\Phi = 10^{-3}$ and $ 2\times 10^{-3}$, respectively.}
\label{fig:plot2} }
\end{figure}

In Fig.~\ref{fig:plot1} we display a logarithmic plot of 
the 
ratio $R \equiv \rho_{\chi}(t)/\rho_{\Phi}(t)$ together with Eq. (\ref{backreaction2}) for 
scenario 1, as defined in Table~\ref{tab1}. 
In order to estimate the sensitivity of $z_{\rm RH} = 1 + m_\Phi \tau$ on $\delta m_\chi^2$, we 
show in Fig.~\ref{fig:plot2} a magnification of the region where $R(z_{\rm RH})=1$. We
also show these plots for the three other scenarios defined in Table~\ref{tab1} (keeping $m_\Phi/m_{\rm pl} = 10^{-6}$ fixed).
\begin{table}[h!]
\begin{tabular}{c||ccc||c|c}
scenario & g & $\lambda$ & $m_\chi^2/m_\Phi^2$ & $\,z_c\,$ & $\kappa_m$\\\hline
$1$ & $4 \times 10^{-4}$ & $10^{-7}$ & $0.1$ & $88$ & $0.8$ \\
$2$ & $6 \times 10^{-4}$ & $10^{-7}$ & $0.4$ & $82$& $0.15$\\
$3$ & $4 \times 10^{-4}$ & $10^{-5}$ & $0.1$ & $47$ & $0.14$\\
$4$ & $6 \times 10^{-4}$ & $10^{-5}$ & $0.4$ & $50$ & $0.06$\\
\end{tabular}
\caption{Definition of the four choices parameter sets. Also shown are the numerical results for $z_c$ and $\kappa_m$, as defined in Eqs~(\ref{backreaction2}) and (\ref{kappadef}). 
\label{tab1}}
\end{table}

We are interested in the change in $\tau$ 
generated by a fluctuation 
$\delta m^2_{\chi} /m^2_{\Phi} \simeq \lambda$. 
Since $\lambda$ is 
tiny, that change can be expressed as 
\begin{equation}
\frac{\delta \tau }{\tau} = \kappa_{m} \frac{\delta m^2_{\chi}}{m^2_\Phi}~~. 
\label{kappadef}
\end{equation} 
From Table~\ref{tab1} we can see that the typical $\kappa_m$ is ${\cal O}(0.1-1)$.

The reader may wonder why we are using larger values 
for $\delta m^2_{\chi}$ that are not consistent 
with the $\lambda$ we choose. 
Since the perturbation 
$\delta \tau$ is linear in $\delta m^2_{\chi}$, the $\kappa_m$ obtained 
this way is unchanged if we were to use smaller values for 
$\delta m^2_{\chi}$.
The reason for this choice of 
$\delta m^2_{\chi}$ is that the plots are easier to 
read. We also repeat that $z_c$ 
was determined with the correct $\lambda$. 
\begin{figure}[t!]
  \centerline{ \mbox{\epsfxsize=9.5truecm \hbox{\epsfbox{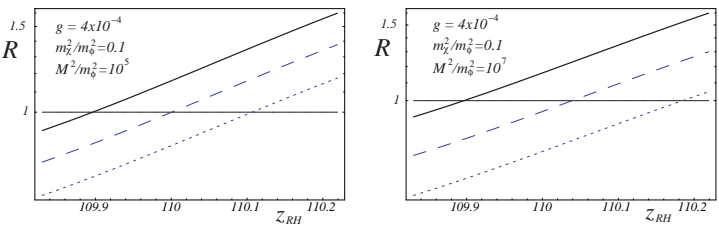}} } }
  {\caption[1]{Logarithmic plot of the 
effect of the fluctuation in the coupling constant $\delta g^2$ on $z_{\rm RH}$. The chosen parameters
are given in each figure. The solid  line corresponds
to $\delta g^2 = 0$, while the long and short dashed lines correspond to
   $\delta g^2/ g^2 = 10^{-3}$ and $2 \times 10^{-3}$, respectively.}
\label{fig:plot3} }
\end{figure}

One may also wonder why the presence of a 
$\delta m^2_{\chi}$ at 
early times has any effect at all, especially given 
that it only persists while $R \lesssim 10^{-5}-10^{-3}$. Parametric resonance 
is dramatic because of stimulated emission. So even if
at earlier times the production of $\chi$ particles 
is affected due to a non-zero $\delta m^2_{\chi}$, this 
will impact the much greater growth occurring at later times. 
A more detailed numerical simulation, including all the effects 
of backreaction and scattering, such as done in \cite{FK} for 
preheating without a fluctuating $\sigma$ field, is needed 
to explore in detail the sensitivity of $\delta \rho /\rho$ to 
super-horizon fluctuations in $\sigma$.

Mathematically, the intuition expressed above 
may be expressed in the following way. 
The density in $\chi$ is approximately 
given by 
\begin{equation} 
\rho_{\chi} \simeq \widetilde{N} \frac{ a^3_0}{a^3} 
{\rm exp}\left[{\int ^{t_{\rm RH}} _{t_0}\!\!\! dt\, \nu(t)}\right] 
\end{equation} 
where $\widetilde{N}$ is a prefactor that depends on the parameters of the model. 
Here $\nu$ is a characteristic exponent leading to exponential growth. 
We approximate 
its dependence on $k$ as given by its value 
near $k \simeq 0$. The coefficient $\nu$ also depends on $m^2_{\chi}$, 
so 
\begin{equation} 
\nu = \nu_0 - \nu_1 \frac{\delta m^2_{\chi}}{m^2_{\Phi}}\Theta(z_c-z) ~.
\end{equation} 
Numerically we find that $\nu_1 / \nu_0 \sim {\cal O}(1)$ and is 
positive. 
A negative correlation is expected, since both the characteristic 
exponents of the Mathieu equation  
and the instability bands  
are the largest near the kinematic limit $A=2q$, corresponding to 
$m_{\chi}\!=\!k\!=\!0$. Increasing $m^2_{\chi}$ removes 
more instability bands from the available phase space. 
Using the approximate formula above, we can solve for 
the change in the reheat time due to a fluctuation $\delta 
m^2_{\chi}$, approximating all the dependence of 
$\delta \tau$ on $\delta 
m^2_{\chi}$ 
as occurring from the exponential. 
This gives 
\begin{equation} 
\frac{\delta \tau}{\tau} \simeq - \frac{\nu_1}{\nu_0} 
\frac{\delta m^2_{\chi}}{m^2_{\Phi}} \frac{z_c}{z_{\rm RH}} 
\simeq \frac{\delta m^2_{\chi}}{m^2_{\Phi}} \frac{z_c}{z_{\rm RH}} ~.
\end{equation} 
This result has ${\cal O}(1)$ agreement with our previous numerical 
computations. (Compare $z_c/z_{\rm RH}$ with $\kappa_m$.) 
It illustrates that $\delta \tau/\tau$ is not 
suppressed by any very small numbers other than 
$\delta m^2_{\chi}/m^2_{\Phi}$. 

We also explore the dependence of $\tau$ on fluctuations 
in $g_{\rm eff}$ \cite{Kofman}. For non-zero particle 
number $n_{\chi}$ the interaction (\ref{geff}) 
introduces a backreaction of $\chi$ on $\sigma$ corresponding 
to an effective mass $m^2_{\sigma,\rm eff} = g^2 \Phi^2 
\langle\chi^2 \rangle/M^2$. 
As before, we cut off the 
fluctuation in $g_{\rm eff} $ 
when $m^2_{\sigma,\rm eff}=3 H^2$. This occurs 
when 
\begin{equation} 
\frac{\rho_{\chi} }{\rho_{\Phi}} = \frac{M^2}{m^2_{\rm pl}} ~.
\label{geffcond}
\end{equation} 
The fluctuation in $g_{\rm eff}$ 
\begin{equation} 
\delta_g \equiv 
\frac{\delta g^2}{g^2}= \frac{H^2_{\rm inf}}{M^2} ~,
\end{equation} 
gives rise to non-Gaussian density perturbations. 

In Fig.~\ref{fig:plot3} we display the ratio $R$ for $g=4 \times 10^{-4}$, 
$m_{\Phi}/m_{\rm pl}=10^{-6}$ and $m^2_{\chi}/m^2_{\Phi}=0.1$. 
We choose two values of $M$ that give $\delta _{g}
=10^{-5}$ 
and $\delta_g =10^{-7}$. According to (\ref{geffcond}), 
the fluctuation in $g_{\rm eff}$ 
is cut off at 
$\rho_{\chi}/\rho_{\Phi}=10^{-7}$ and $10^{-5}$, respectively, 
corresponding to $z_c=26$ and $z_c=47$. For both of these parameters 
we find that there is a large linear effect which we express as 
\begin{equation} 
\frac{\delta \tau }{\tau} = \kappa_{g} \delta_g~.
\end{equation} 
For $\delta_g=10^{-5}$ we find  $\kappa_{g}=0.9$ and 
for $\delta_g=10^{-7}$ we find  $\kappa_{g}=1.4$. 
As in the previous case, in obtaining our plots we used 
larger values of $\delta_g$ to determine 
$\kappa_g$.

In conclusion, we have shown that during 
preheating, 
interactions of the ``decay products'' 
of the 
inflaton with other 
light scalar fields can give rise to super-horizon
mass fluctuations in 
these decay products. 
These fluctuations will then
 give rise to density perturbations of the universe. Depending on 
the coupling
 of the decay products of the 
inflaton to the light scalar fields, the dominant 
 density perturbations generated 
from this effect will be either Gaussian or non-Gaussian.

This work was supported by the 
Department of Energy. 
under the contract DE-FG03-92ER40701.

\end{document}